\pdfoutput=1
\documentclass[conference]{IEEEtran}
\IEEEoverridecommandlockouts

\usepackage[utf8]{inputenc}
\usepackage[T1]{fontenc}
\usepackage[american]{babel}
\usepackage{graphicx}
\usepackage{listings}
\usepackage{float}
\usepackage{amsmath,amssymb,exscale}
\usepackage{blindtext, graphicx}
\usepackage{verbatim}
\usepackage{algorithm}
\usepackage{algpseudocode}
\usepackage{fancyvrb}
\usepackage{bera}
\usepackage{amsmath}
\usepackage{mathtools}
\usepackage{lipsum}
\usepackage{subfigure}
\usepackage{cite}
\usepackage{scalerel}
\usepackage{tikz}

\usepackage[bookmarks=false]{hyperref}

\def\BibTeX{{\rm B\kern-.05em{\sc i\kern-.025em b}\kern-.08em
    T\kern-.1667em\lower.7ex\hbox{E}\kern-.125emX}}

\makeatletter 
\let\old@ps@headings\ps@headings 
\let\old@ps@IEEEtitlepagestyle\ps@IEEEtitlepagestyle 
\def\confheader#1{%
\def\ps@headings{%
\old@ps@headings%
\def\@oddhead{\strut\hfill#1\hfill\strut}%
\def\@evenhead{\strut\hfill#1\hfill\strut}%
}%
\def\ps@IEEEtitlepagestyle{%
\old@ps@IEEEtitlepagestyle%
\def\@oddhead{\strut\hfill#1\hfill\strut}%
\def\@evenhead{\strut\hfill#1\hfill\strut}%
}%
\ps@headings%
} 
\makeatother 

\confheader{%
16th IEEE INTERNATIONAL WORKSHOP ON INFORMATION FORENSICS AND SECURITY (WIFS) 2024
} 

\newcommand\copyrighttext{%
  \footnotesize \textcopyright 2024 IEEE. Personal use of this material is permitted.
  Permission from IEEE must be obtained for all other uses, in any current or future 
  media, including reprinting/republishing this material for advertising or promotional 
  purposes, creating new collective works, for resale or redistribution to servers or 
  lists, or reuse of any copyrighted component of this work in other works. 
  DOI: }
\newcommand\copyrightnotice{%
\begin{tikzpicture}[remember picture,overlay]
\node[anchor=south,yshift=10pt] at (current page.south) {\fbox{\parbox{\dimexpr\textwidth-\fboxsep-\fboxrule\relax}{\copyrighttext}}};
\end{tikzpicture}%
}
\begin{document}

\title{Vulnerabilities in Machine Learning-Based Voice Disorder Detection Systems}

\author{\IEEEauthorblockN{Gianpaolo Perelli, Andrea Panzino, Roberto Casula, Marco Micheletto, Giulia Orrù, Gian Luca Marcialis}
\IEEEauthorblockA{Department
of Electrical and Electronic Engineering, University of Cagliari, Italy\\
Email: {\{gianpaolo.perelli, andrea.panzino, roberto.casula, marco.micheletto, giulia.orru, marcialis\}@unica.it,}}

}

\IEEEoverridecommandlockouts 
\IEEEpubid{\makebox[\columnwidth]{\copyrightnotice } 
\hspace{\columnsep}\makebox[\columnwidth]{ }} 

\maketitle

\begin{abstract}

The impact of voice disorders is becoming more widely acknowledged as a public health issue. Several machine learning-based classifiers with the potential to identify disorders have been used in recent studies to differentiate between normal and pathological voices and sounds. In this paper, we focus on analyzing the vulnerabilities of these systems by exploring the possibility of attacks that can reverse classification and compromise their reliability. Given the critical nature of personal health information, understanding which types of attacks are effective is a necessary first step toward improving the security of such systems.
Starting from the original audios, we implement various attack methods, including adversarial, evasion, and pitching techniques, and evaluate how state-of-the-art disorder detection models respond to them. Our findings identify the most effective attack strategies, underscoring the need to address these vulnerabilities in machine-learning systems used in the healthcare domain.

\end{abstract}


\section{Introduction}
\label{sec:intro}

Voice disorders manifest as variations from normal voice quality, pitch, and loudness concerning an individual's age, gender, and cultural background \cite{boone2005voice}. These disorders impact a substantial portion of the population, with estimates suggesting that up to 20\% of people may encounter a voice disorder during their lifetime \cite{huston2024prevalence}. Conditions can vary from minor discomfort to severe dysphonia and span functional to malignant categories. The diagnosis of voice disorders usually encompasses a comprehensive clinical examination, including interviews, auditory-perceptual judgments, acoustic analysis, and laryngoscopy, with biopsy required in cases suspecting malignancy \cite{umeno2020summary}.
Given these disorders' wide range and prevalence, the requisite in-depth diagnostic procedures demand considerable time from patients and clinicians and lead to significant economic burdens on healthcare systems \cite{cohen2012direct}. Moreover, the complexity and time-intensiveness of these processes can result in delayed diagnoses, potentially exacerbating the patient's condition and further complicating treatment \cite{paul2013diagnostic}.

In this context, machine learning (ML), particularly through deep neural networks, offers a promising advancement in this area by potentially enhancing both the speed and accuracy of diagnosis. The application of ML in healthcare has led to significant developments in computer-aided diagnosis systems, which assist doctors in making diagnoses through computer-generated outputs \cite{alanazi2022using}.
In particular, such models are used in medicine to interpret imaging data (like MRI and CT scans) for detecting cancers, fractures, and other abnormalities, to monitor real-time patients, predicting critical events like sepsis or heart failure \cite{alotaibi2019implementation} and in many other areas.
Among these diverse applications, voice disorder detectors have shown remarkable potential, leveraging acoustic features extracted from voice recordings to categorize them into normal or pathological \cite{gupta2024voice}.
However, the integration of such technology introduces new vulnerabilities. Particularly concerning is the potential for adversarial manipulation, a concept well-documented in other ML domains, such as image and speech recognition \cite{carlini2018audio}.  

Adversarial attacks involve deliberately altering input data to deceive the model into erroneous predictions \cite{xu2020adversarial}. These manipulations are often imperceptible or seemingly benign to human listeners, yet they can drastically alter the model's output.
An especially concerning application of such attacks could be the creation of manipulated audio recordings designed to falsely suggest the presence of a medical condition in a target individual. Such deceptive practices could have profound implications: a person might be unjustly disqualified from employment opportunities or be subjected to inflated health insurance rates based on fabricated evidence of illness, and a public figure could suffer damage to their reputation. For instance, fake pathological recordings could be used in cyberbullying or harassment campaigns to manipulate public perception of individuals.
This situation is reminiscent of the challenges posed by deepfake technologies \cite{zhang2022deepfake}, where the authenticity of visual and audio content can be convincingly altered, leading to the spread of misinformation and potential harm before any form of expert validation can occur. 

In this context, our paper explores adversarial attacks and audio manipulations within the context of voice disorder detection (VDD). By identifying and analyzing the potential threats posed by adversarial or manual manipulations, we aim to highlight the critical need for robust countermeasures. Ensuring the security and integrity of disorder detection systems against manipulations is paramount in safeguarding the diagnoses' accuracy and trust in these diagnostic tools.

The paper is organized as follows. Section \ref{sec:relatedwork} reviews the current literature on VDD and audio adversarial attacks and manipulations. Section \ref{sec:attacks} describes the attacks implemented to assess the robustness of voice disorder detection systems. Section \ref{sec:protocol} describes the protocol used to conduct our evaluation, while section \ref{sec:results} reports the obtained results. Finally, conclusions are drawn in Section \ref{sec:conclusions}.

\label{sec:relatedwork}
\subsection{Voice Disorder Detection}
In recent years, the field of voice disorder detection has made significant progress, with the development of various methodologies aimed at more accurately distinguishing between healthy and pathological voices. Central to these advancements is applying ML algorithms to vocal data analysis. These algorithms utilize a range of classifiers and leverage key acoustic features, such as Mel-Frequency Cepstral Coefficients (MFCC) \cite{ eskidere2015voice} or Multidimensional Voice Program (MDVP) \cite{al2017investigation} to enhance diagnostic precision.
Among the ML techniques, SVM has seen widespread use due to its effectiveness in classifying pathological and healthy voices. Studies employing SVM with MFCC have reported significant accuracies, though often limited by small datasets \cite{idrisoglu2023applied}. 
For instance, Al-Dhief et al. \cite{al2021voice} utilized SVM with MFCC methodology, achieving an 91.17\% accuracy using voice signals from the Saarbrücken voice database (SVD)\footnote{https://stimmdb.coli.uni-saarland.de/}. Souissi et al. \cite{souissi2015dimensionality} enhanced the SVM training process by integrating MFCC with Linear Discriminant Analysis (LDA) for more efficient dimensionality reduction, leading to an 86.44\% accuracy rate in detecting voice pathologies within a subset of the SVD. 
Other common classifiers used in the literature include artificial neural networks \cite{teixeira2017vocal}, hidden Markov models  \cite{benhammoud2018automatic}, and Gaussian mixture models  \cite{ali2017automatic}. 

Parallel to these efforts, advancements in end-to-end ML and deep learning techniques have further expanded the possibilities for VDD. 
Studies utilizing deep learning have mainly considered Convolutional Neural Network (CNN) models to extract acoustic features from spectrogram-based voice data automatically \cite{reid2022development, powell2019decoding} or in combination with hand-crafted features \cite{fang2019detection}. Nevertheless, the efficacy of these models can be compromised by limited dataset sizes, which often lead to overfitting and diminished performance on broader datasets. 

\subsection{Audio Adversarial attacks}

Similar to the image domain \cite{khamaiseh2022adversarial}, adversarial attacks have also exposed considerable vulnerabilities in audio-based ML systems such as Automatic Speech Recognition (ASR) and Automatic Speaker Verification (ASV). These attacks cleverly embed perturbations into audio signals to deceive detection models, ensuring the alterations remain imperceptible to human listeners. 
Techniques involving the manipulation of waveforms, spectrograms, or MFCC features have proven effective in causing ASR systems to interpret spoken language erroneously \cite{carlini2018audio}. Similarly, adversarial examples have been crafted to mislead ASV systems into falsely identifying a voice as belonging to a particular individual \cite{liu2019adversarial}.
In general, adversarial attacks can be broadly categorized based on the attacker's knowledge of the target model. In a white-box attack scenario, the attacker possesses complete knowledge of the model's architecture and parameters, enabling precise and potent adversarial example crafting \cite{szegedy2013intriguing}. In this scenario, gradient-based methods, such as the Fast Gradient Sign Method (FGSM) and Projected Gradient Descent (PGD), leverage the model's gradient information to create perturbations that maximize prediction error while remaining imperceptible to human ears \cite{madry2019deep}. In contrast, black-box attacks assume that the attacker has no direct access to the target model information but can observe its output given specific inputs, making these attacks more reflective of real-world scenarios \cite{papernot2017practical}. Several works exploit this strategy to fool audio-based systems \cite{ge2023advddos, zheng2021black}.

Despite the extensive exploration of adversarial attacks within the broader audio domain, the application of such attacks on voice disorder detection systems has, to the best of our knowledge, not yet been reported. Given the critical nature of healthcare applications, the potential implications of such adversarial interventions are manifold and particularly concerning. Firstly, introducing adversarial perturbations could lead to erroneous diagnosis, either by masking the presence of a disorder or by falsely indicating one, thereby undermining the reliability of these systems in clinical settings. Furthermore, the integrity of patient data could be compromised, leading to privacy violations and loss of trust in digital healthcare solutions. The robustness of voice disorder detection models must be severely tested, exposing vulnerabilities that could be exploited to degrade system performance over time.

\section{Adversarial attacks for disorder detectors}
\label{sec:attacks}
\begin{figure}
    \centering
    \includegraphics[width=0.27\textwidth]{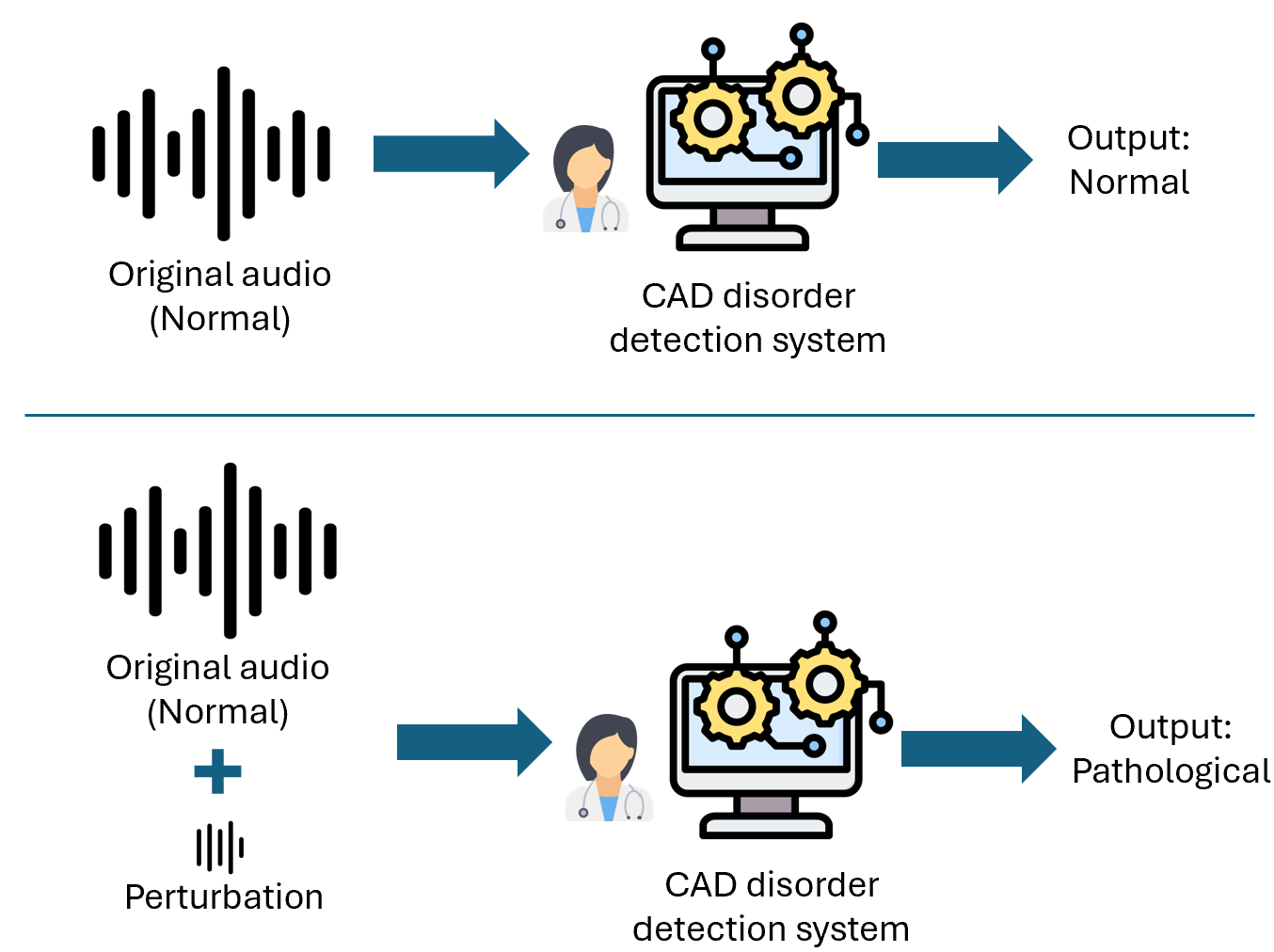}
    \caption{High-level diagram of attacks for voice disorder detection systems.}
    \label{fig:geomdistr}
\end{figure}
Integrating ML into the diagnostic process offers an innovative solution that optimizes diagnostic times and precision. Computer-aided diagnosis systems have already shown promise in image processing for disease prediction \cite{patil2023review} and can thus be applied to voice disorder detection, providing new methods to analyze acoustic data. However, the advancement of these technologies introduces potential risks that need careful examination. To address these concerns, this study examines the vulnerabilities of voice disorder detection systems to targeted manipulations used for side-channel attacks (Fig. \ref{fig:geomdistr}). We investigate how such attacks can deliberately alter the system's output, causing it to erroneously classify 'normal' voice samples as 'pathological'. To assess different levels of vulnerability, we implement both simple attacks involving tone and pitch manipulation and more complex attacks, such as adversarial ones.

For this purpose, we utilize a comprehensive approach encompassing white-box and black-box attack scenarios. The white-box attacks are conducted by leveraging two adversarial techniques based on gradients, exploiting complete visibility of the model's architecture and parameters. 
In parallel, black-box attacks are emulated by altering the pitch or adding an extraneous but imperceptible tone to the audio samples. This approach allows for a more realistic attack scenario, where the adversary lacks direct knowledge of the model's inner workings and relies solely on its output to provoke misclassification.
The proposed two-faceted approach ensures accurate and thorough investigations into the robustness of VDD systems against various adversarial tactics.

\subsubsection{White-box attacks} 
In this study, we applied two well-known gradient-based techniques, namely Fast Gradient Sign Method (FGSM) and Projected Gradient Descent (PGD), to explore the vulnerability of ML models to white-box adversarial attacks. The FGSM approach manipulates the input data by calculating the loss gradient with respect to the input features, identifying the direction in which slight changes can most effectively mislead the model. This gradient sign is then scaled by a value called epsilon ($\epsilon$), which controls the intensity of the perturbation added to the original input. The PGD approach can be considered an extension of the FGSM technique: it iteratively applies multiple small gradient updates to perturb the input data.  Again the perturbation is adjusted within a predefined $\epsilon$ boundary, but iteratively to find the optimal direction and magnitude.
\\
Since the architectures and parameters of the models must be known by the attacker for precise calculation of gradients, we define such attacks as white-box.
\subsubsection{Black-box attacks}
These attacks are based on the principle that the attacker has no knowledge of the model's internal workings, including its architecture, weights, or the specific algorithms it employs. Instead, the attacker only has access to the model's input-output behavior. In this study, we applied two particular black-box evasion attacks: tone evasion attack and audio pitching.
\\
The first is based on inserting a specific tone into an audio sample. In this regard, it is possible to modify the tone parameters to be inserted (e.g., in terms of amplitude, frequency, and phase) to balance the perturbation induced on the audio sample with the auditory perception of the disturbance.
In our analysis, we selected several sine waves with varying frequencies and amplitudes. Specifically, we chose frequencies of 50, 75, 100, 125, and 150 Hz. For each frequency, we considered five different amplitudes to evaluate their auditory influence on the original audio: from causing a slight effect (0.2, 0.3, 0.4) to inducing a significant auditory disturbance (0.8, 0.9). Notably, higher amplitude values increase both the perturbation on the audio samples and the auditory perception of the disturbance.
\\
A second attack proposed in this work is based on the substantial modification of the pitch of the audio samples. This experiment aims to determine if a voice disorder detection system is resilient enough to pitch modification on test samples. In this case, all the experiments were carried out by a pitch down of 5 steps (on a total of 12 per octave). This particular value is chosen to slightly modify the samples but leave all auditory properties as close to the original audio.

\section{Experimental protocol}
\label{sec:protocol}
\subsection{Datasets}

For the experimental analysis, we used two different datasets, the HUPA dataset \cite{app12168129} and the Saarbruecken Voice Database (SVD).
The HUPA dataset recorded by
Universidad Politécnica de Madrid and Príncipe de Asturias Hospital of Alcalá de Henares is composed of two types of audio: 100 \textit{normal} samples are related to users who do not have vocal difficulties, while 100 \textit{pathol} samples are related to users with voice disorders (vocal fold polyps, nodules, edema, vocal leukoplakia, etc.). All these audios have a length between 1 and 3 seconds and a 25 kHz sampling rate. The SVD database is an exhaustive collection of voice recordings by more than 2000 persons. It was collected at the Institute of Phonetics and Phoniatry, Caritas Clinic St. Theresia, Saarbrücken. The dataset comprises audio samples of the sustained vowel sounds /a/, /i/, and /u/ uttered at different intensities: normal, high, and low, along with variations in a rising and falling pitch pattern. Each sample was recorded at a frequency of 50 kHz. For the purposes of this study and for the sake of space, we focused exclusively on the vowels /a/ uttered at normal pitch. We then selected three prevalent voice disorders: vocal fold cyst, vocal fold polyp, and unilateral vocal fold paralysis, to ensure a comparable sample size with the HUPA datasets. After this process, we obtained a subset of 520 recordings from individuals aged 16 and older, balanced with 260 normal and 260 pathological samples.


\subsection{Voice disorder detectors}
For the implementation of voice disorder detectors, two different types of features were analyzed: (i) a low-level acoustic, mel-spectrogram feature, calculated as one spectrogram at mel-frequency from the Fourier spectrum using a nonlinear transformation at the frequency axis and normalized between 0 and 1; (ii) the Mel Frequency Cepstral Coeficients (MFCC) feature, obtained by applying Discrete Cosine Transform (DCT) to convert log Mel spectrum in the time domain.

Four different classifiers were subsequently used:
\begin{itemize}
    \item The simple CNN proposed in \cite{guan2019evaluation}, composed by two transposed 2D convolutional layers, a pooling layer, and multiple fully-connected layers.
    \item A CNN feature-extractor followed by an SVM classifier (linear kernel), as proposed in \cite{guan2019evaluation}. 
    \item  MobileNetV3Small \cite{howard2019searching, yousif2023generic}, a deep but at the same time fast network with low computational complexity. In particular, we used a model pre-trained on natural images. To adapt the network and carry out fine-tuning, we replaced the last layer with a dense two-neuron layer and modified the first layer to manage a single-channel input. We used stochastic gradient descent as optimizer.
    \item MobileNetV3Small followed by an SVM classifier: in this case, the last dense layer of the MobileNet has been replaced with a Radial Basis Function SVM kernel.
\end{itemize}
From the combination of these feature extractors and classifiers, we obtained six different voice detectors that we used to evaluate the danger of the attacks created. 
To select the models on which to carry out the attack, each dataset was divided into training, validation and test (70\%, 10 \%, 20 \%), and after a 5-fold cross-validation, the best model was selected.

Before extracting the features, each sample was subjected to a pre-processing phase to adapt to the input of the specific network. For MobileNet-based detectors, each audio file is divided into multiple 200 ms snippets with an overlap of 160 ms (sampled at 25 kHz). For CNN-based detectors, each audio file is divided into multiple 1 s snippets with an overlap of 900 ms (sampled at 16 kHz). \\

Since each snippet represents a sample for the implemented models, the experimental evaluation produced both an accuracy on the snippets and on the entire audio file.  The classification of the entire audio file was achieved by majority voting on the snippet predictions.
In particular, in the white-box scenario, since the attacker has knowledge of the entire architecture, he/she can attack the single correctly classified snippet. In the black box case, in order to obtain comparable results with the white box, we distinguish two scenarios: the attacker is completely uninformed, so he/she attacks the entire audio file, or the attacker is aware of the division into snippets and can thus attack a single snippet. Since the attacks were only carried out on correctly classified healthy samples, results are reported in terms of True Positive Rate (TPR).


\section{Experimental Results}
\label{sec:results}
\subsection{Black-box evaluation}

\begin{figure}
    \centering
    \includegraphics[width=8.5cm]{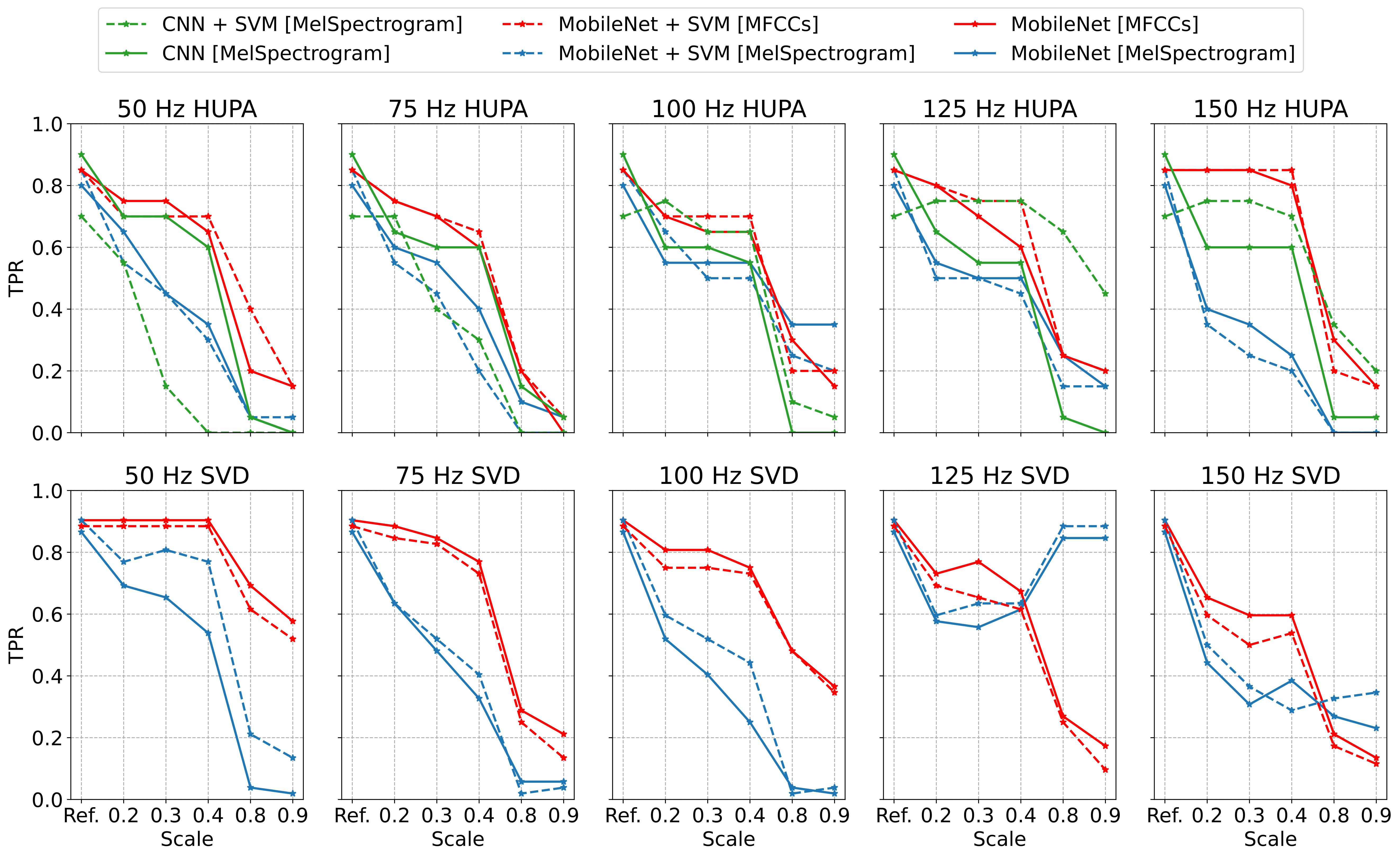}
    \caption{Results of tone-based evasion attacks.}
    \label{fig:acc_freq}
\end{figure}

\begin{figure}
    \centering
    \includegraphics[width=7.5cm]{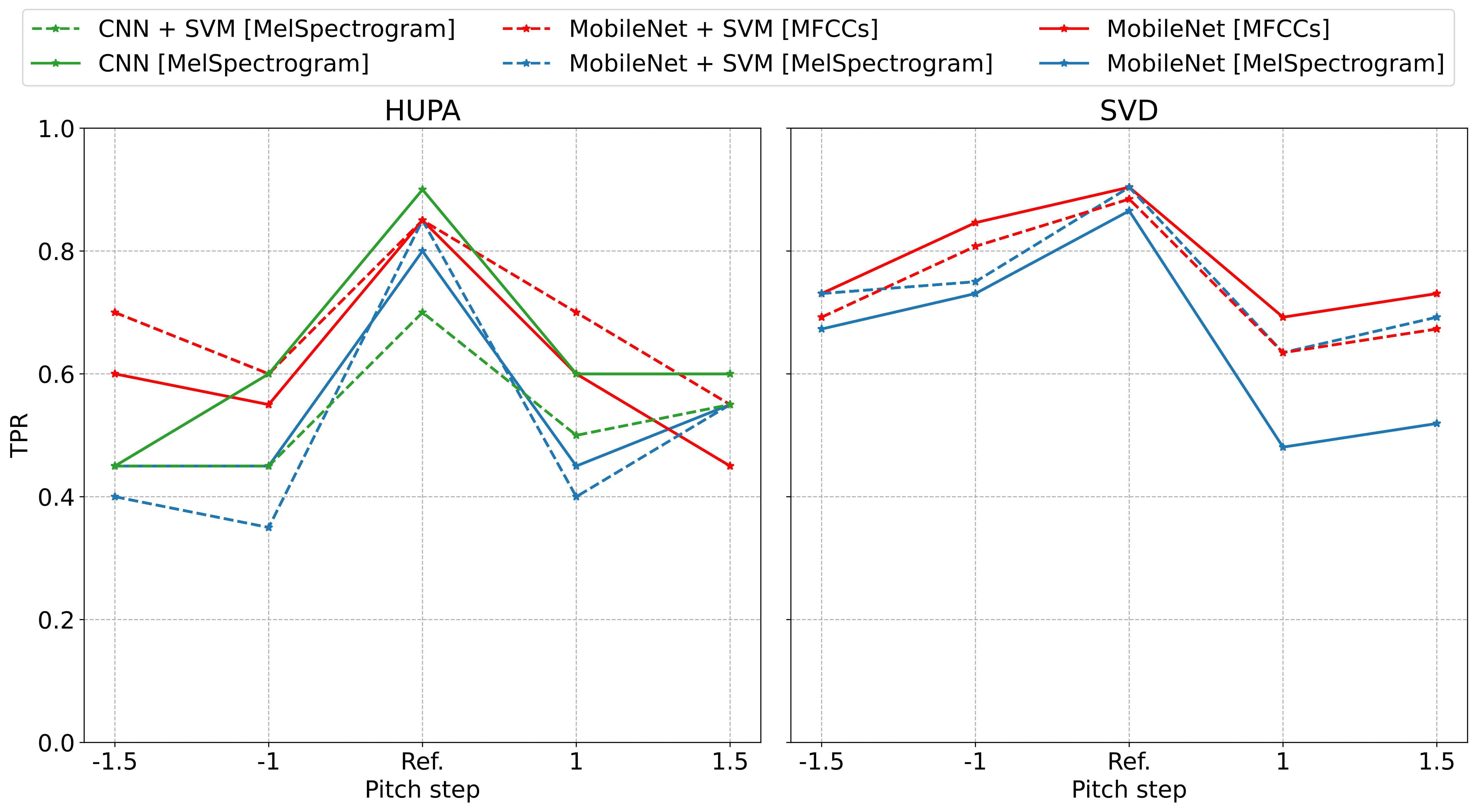}
    \caption{Results of the pitch-based evasion attacks.}
    \label{fig:acc_pitch}
\end{figure}

The results of the black-box evaluation for tone and pitch evasion attacks are reported in Fig.s \ref{fig:acc_freq} and \ref{fig:acc_pitch}, respectively. Evaluations were carried out using tones at different frequencies and scales. The tone-based evasion attack results (Fig. \ref{fig:acc_freq}) show that adding a tone generally reduces the accuracy of the model evaluated. In particular, it can be seen that the deterioration in accuracy is mainly related to the amplitude scale used for the attack. Higher tone scales result in more significant errors in classification: in particular, this manipulation causes the TPR, which originally ranges between 70\% and 90\%, to drop to values below 40\%. This is reasonable since higher-scale tones introduce more noise into the samples, leading to greater classification errors. On the other hand, by fixing the scale and varying the frequency from $50$ to $150$ Hz, less influence of the latter was seen in terms of classifier accuracy. Additionally, the results also show that overall the MFCC feature is more robust to this kind of manipulation.

Fig. \ref{fig:acc_pitch} shows the results of the pitch-based evasion attacks. Changing the pitch in all cases leads to a decrease in TPR. However, it is also evident that the drop in performance is not as high as that seen in the case of tone-based attacks. Furthermore, except for using a step of $-1$, the drop in performance generally seems uncorrelated with the pitch step values used. The findings generally highlight the non-robustness of the models analyzed to normal variations in tone and pitch, which may also occur not necessarily for attacks. In fact, although the mel-spectrogram features are more sensitive to these variations, all the combinations of classifiers/features analyzed report a significant decrease in TPR.

\subsection{White-box evaluation}

\begin{figure}
    \centering
    \includegraphics[width=8cm]{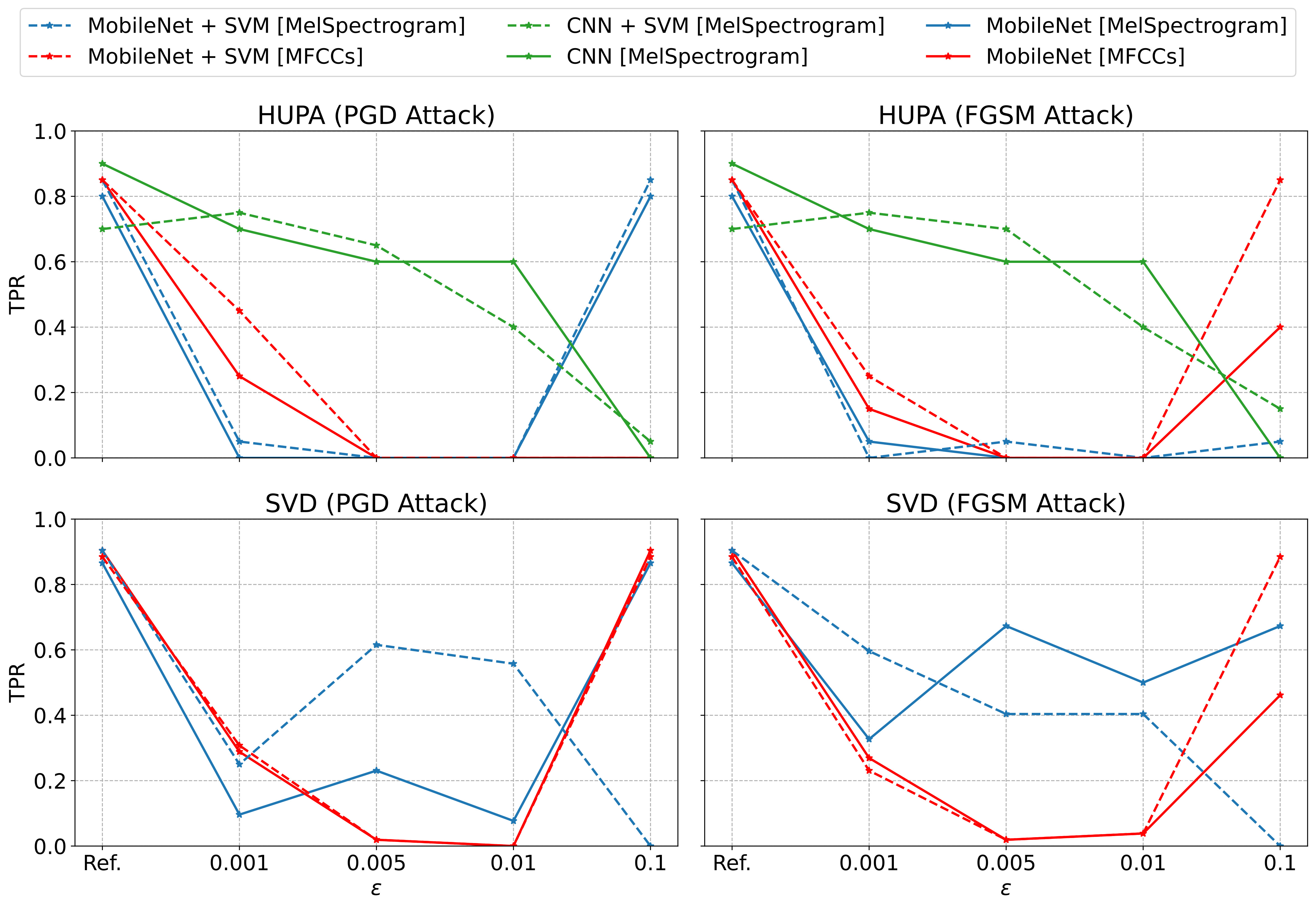}
    \caption{Results of PGD and FGSM attacks, using different $\varepsilon$ values.}
    \label{fig:whitebox}
\end{figure}

In the white-box scenario (Fig. \ref{fig:whitebox}), both \textit{PGD} and \textit{FGSM} adversarial attacks demonstrated their effectiveness in deceiving the classifiers even with minimal perturbations ($\epsilon=0.001$), especially when using a MobileNet architecture. This vulnerability is even more evident at higher epsilon values ($\epsilon=0.1$), where the perturbation noise is severe enough to flatten the classifier's scores, either on the normal classification (FGSM) or on the pathological classification (PGD). On the other hand, CNNs show greater robustness against such attacks, maintaining better performance and stability across various perturbation levels.

\begin{figure*}
    \centering
    \subfigure{\includegraphics[width=0.31\linewidth]{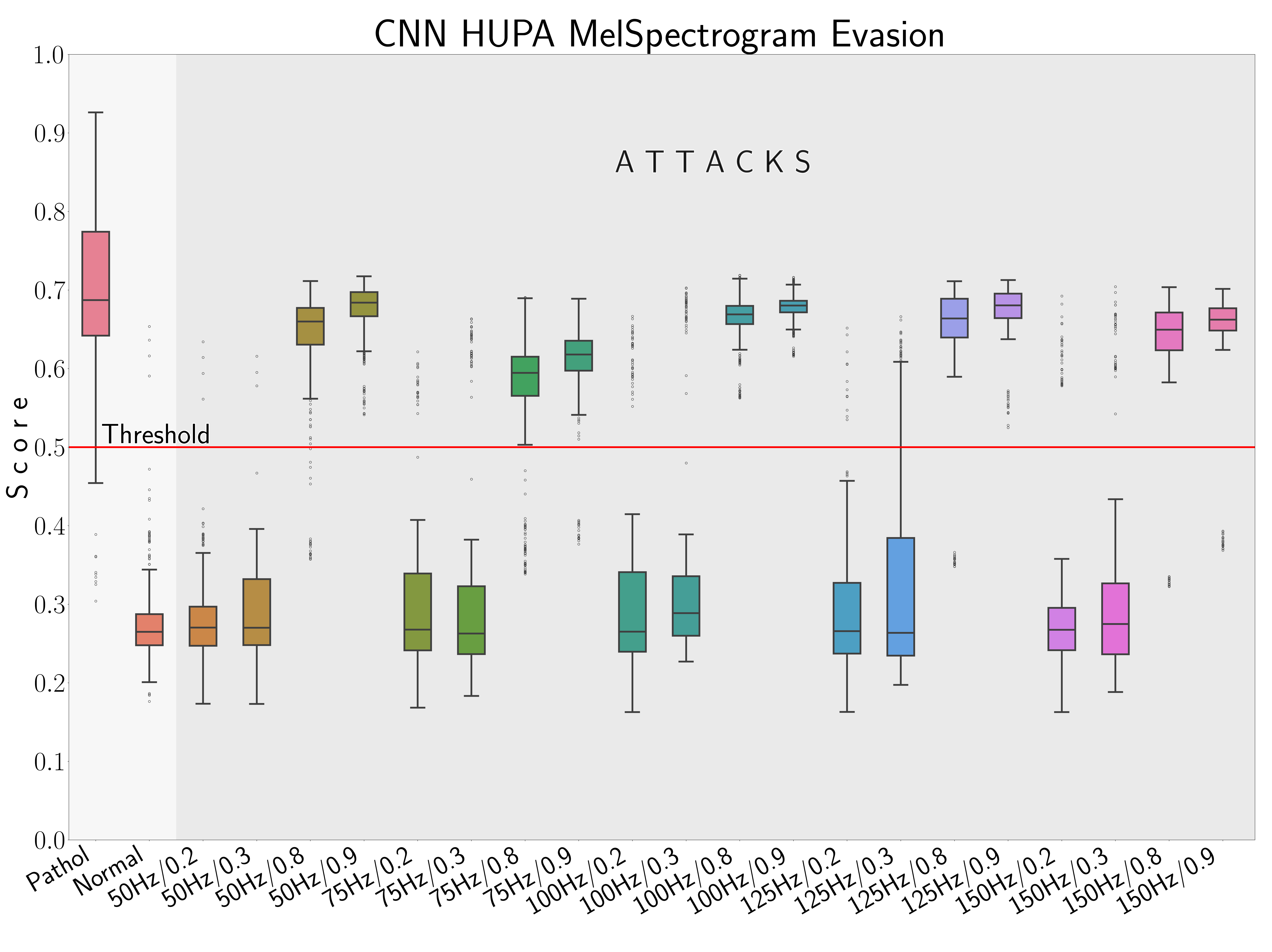}}
    \subfigure{\includegraphics[width=0.31\linewidth]{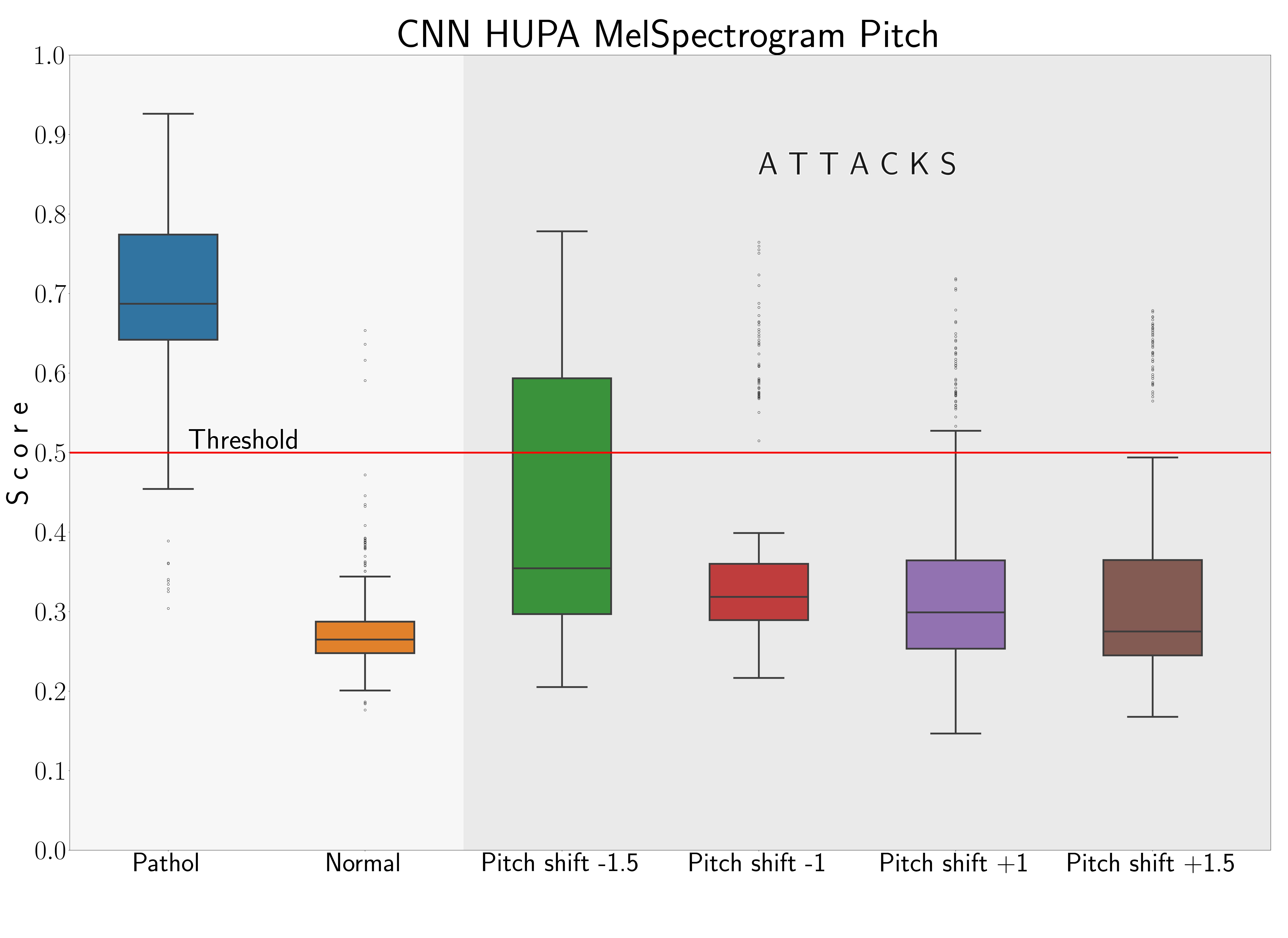}}
        \subfigure{\includegraphics[width=0.31\linewidth]{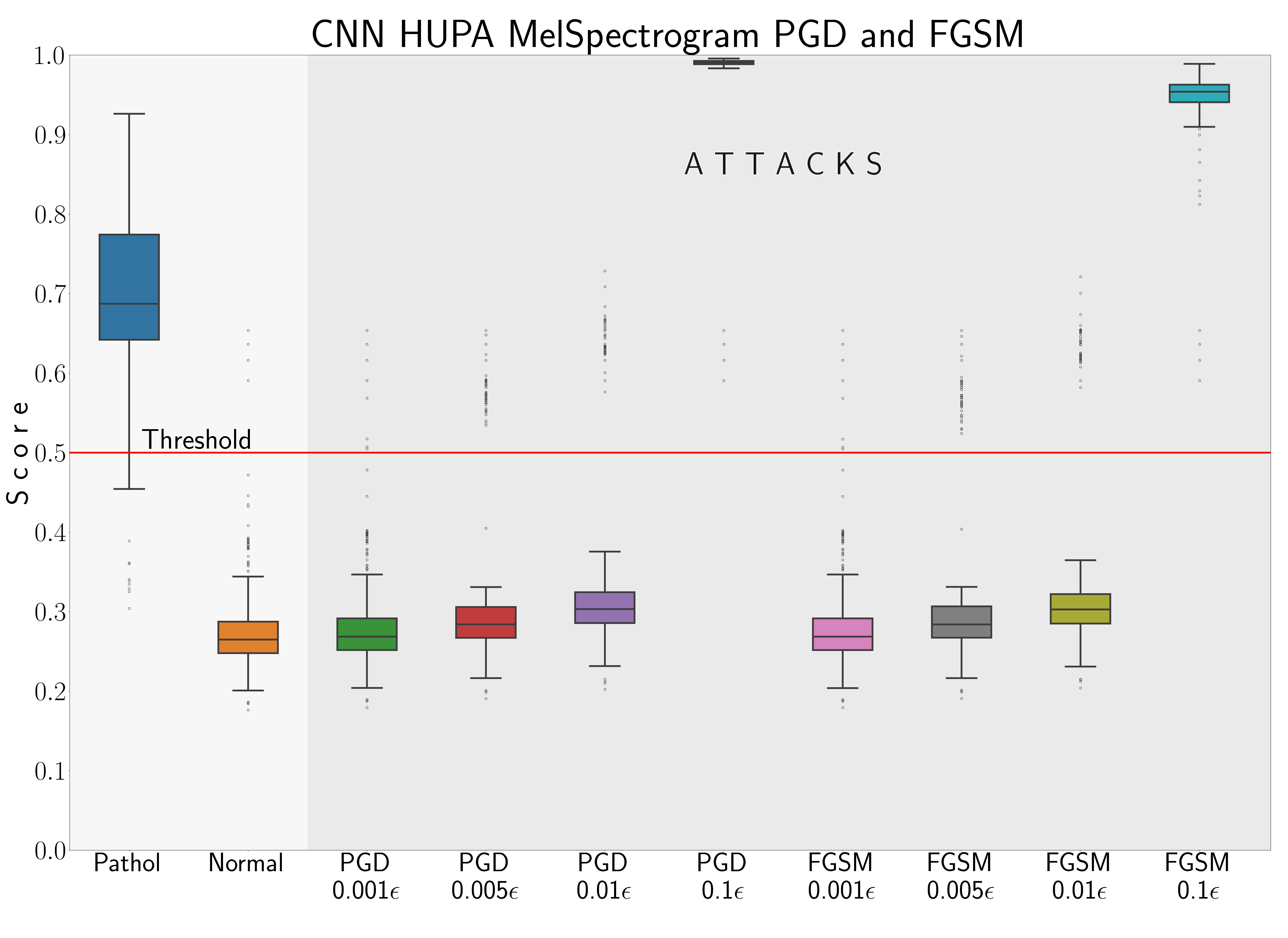}}

        \subfigure{\includegraphics[width=0.31\linewidth]{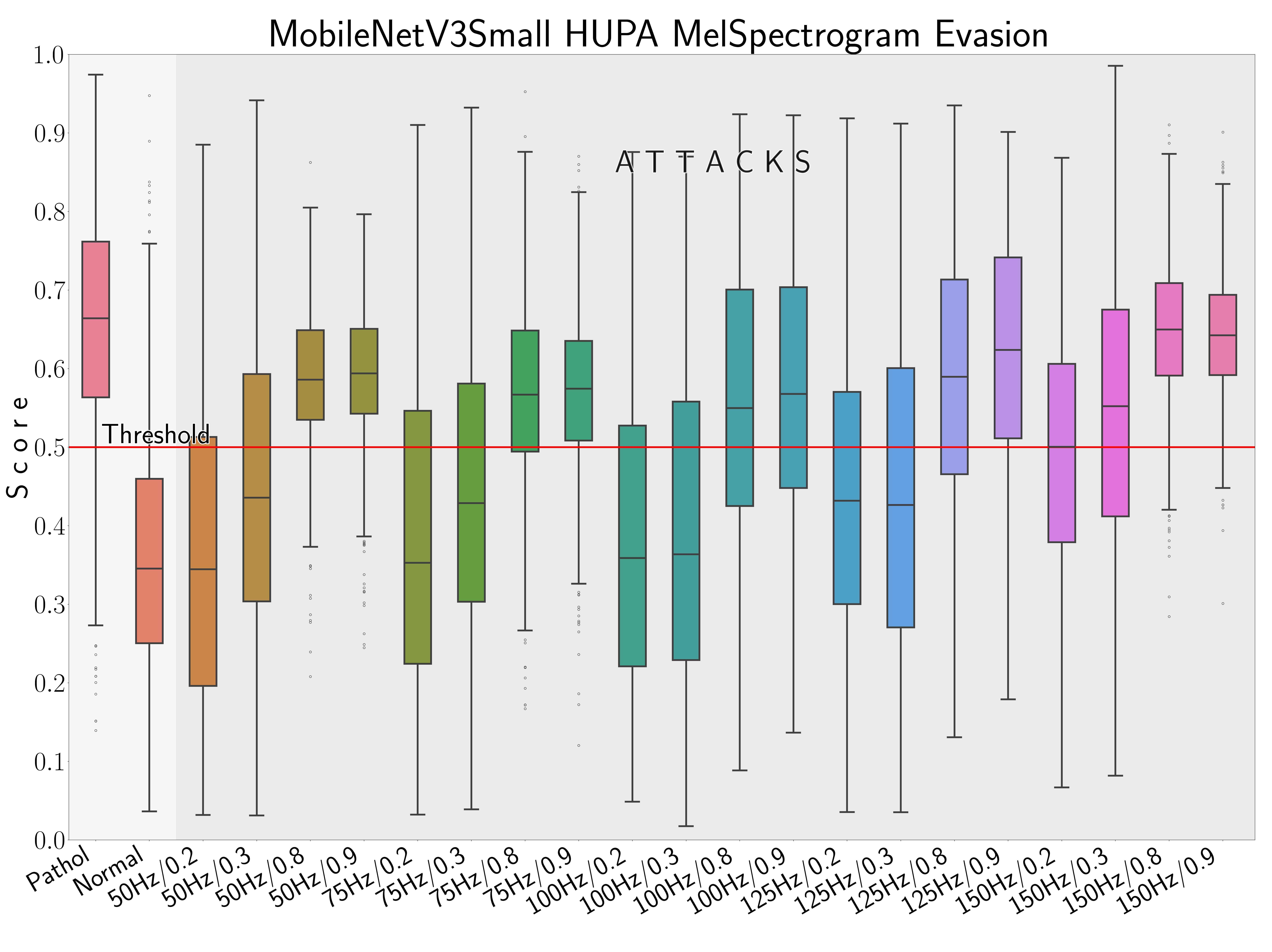}}
    \subfigure{\includegraphics[width=0.31\linewidth]{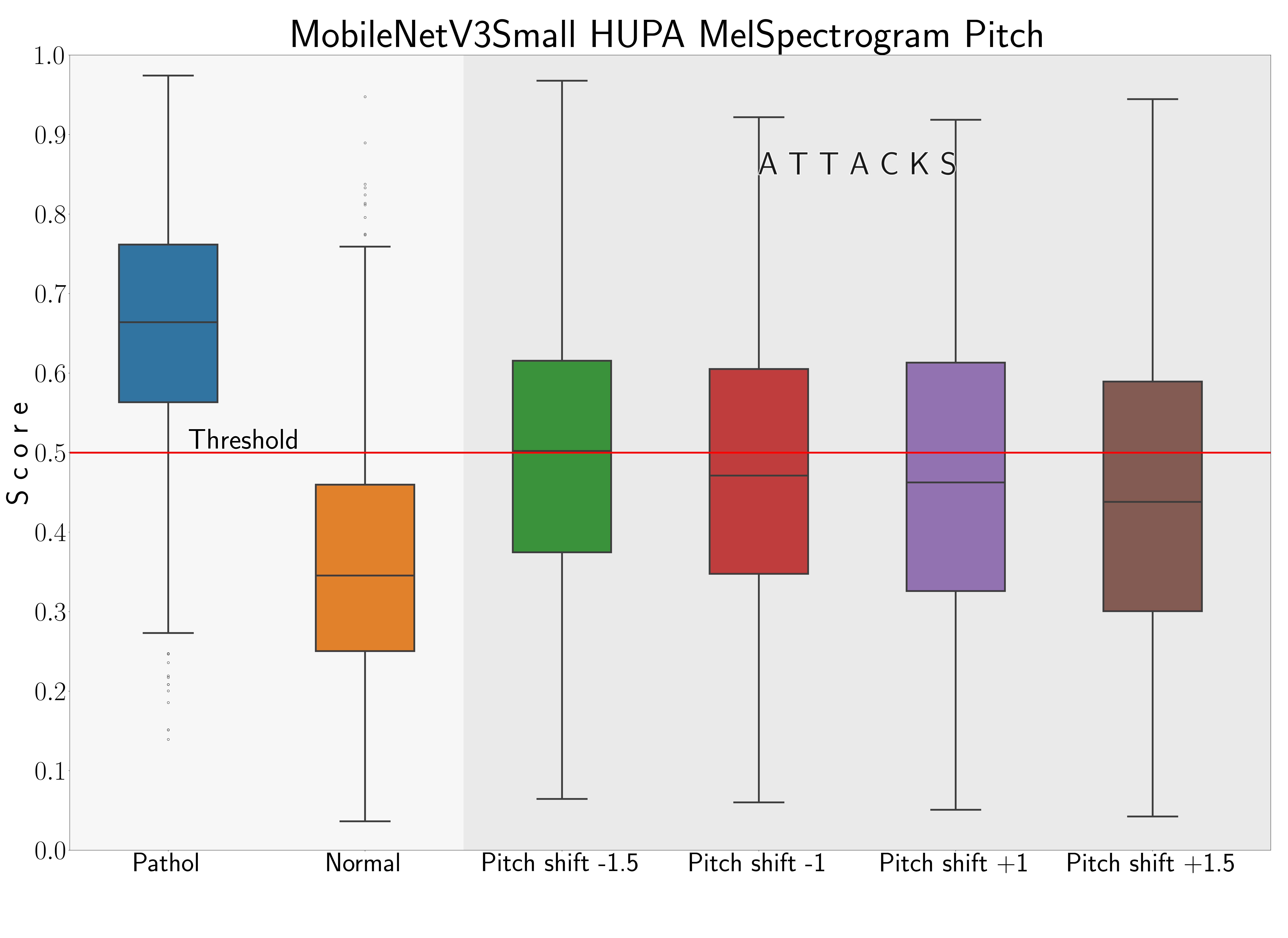}}
        \subfigure{\includegraphics[width=0.31\linewidth]{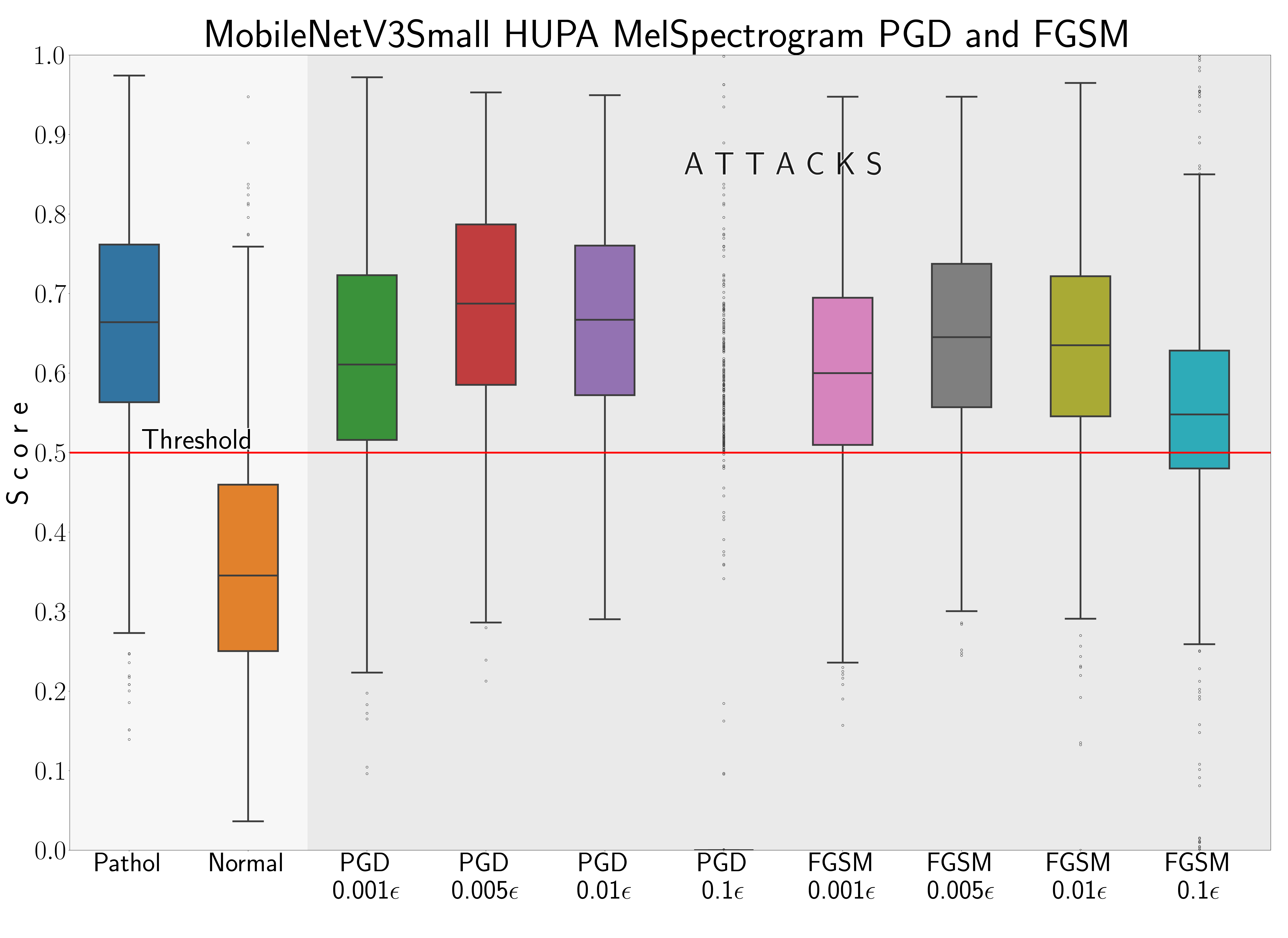}}
    \caption{Boxplots of scores for snippets of correctly classified files obtained with mel-spectrogram-based classifiers on the HUPA dataset. The boxplots compare the scores from the original unperturbed snippets (pathol and normal) with those subjected to black-box and white-box attacks.}
    \label{fig:boxplot}
\end{figure*}

It is interesting to note the effectiveness of the \textit{FGSM} attack, which, being much faster than \textit{PGD}, can be used in real-time, deceiving the system without causing any slowdown.

As a final analysis, we computed the boxplots in Fig. \ref{fig:boxplot} to illustrate the distribution of classifier scores for snippets of correctly classified files before and after attacks. For the sake of space, only the results of the classifiers based on the mel-spectrogram features on the HUPA dataset have been reported. However, they are representative of the overall trends observed. 
Each boxplot displays several key statistics: the median, the interquartile range (IQR, the box length which represents the spread of the middle 50\% of the data), and potential outliers (points beyond the whiskers, which extend to 1.5 times the IQR from the quartiles). Overall, the original snippets exhibit lower median scores and narrower IQRs, indicating consistent and reliable classifier performance on clean data. In contrast, the attacked snippets show higher median scores and wider IQRs, highlighting the significant impact of perturbations on classification accuracy. Additionally, the presence of outliers in the attacked data suggests that the perturbations are causing extreme variations in the classifier's scores, further emphasizing the systems' vulnerability to adversarial attacks.

\section{Discussion and conclusion}
\label{sec:conclusions}

\begin{figure}
    \centering
    \includegraphics[width=8.5cm]{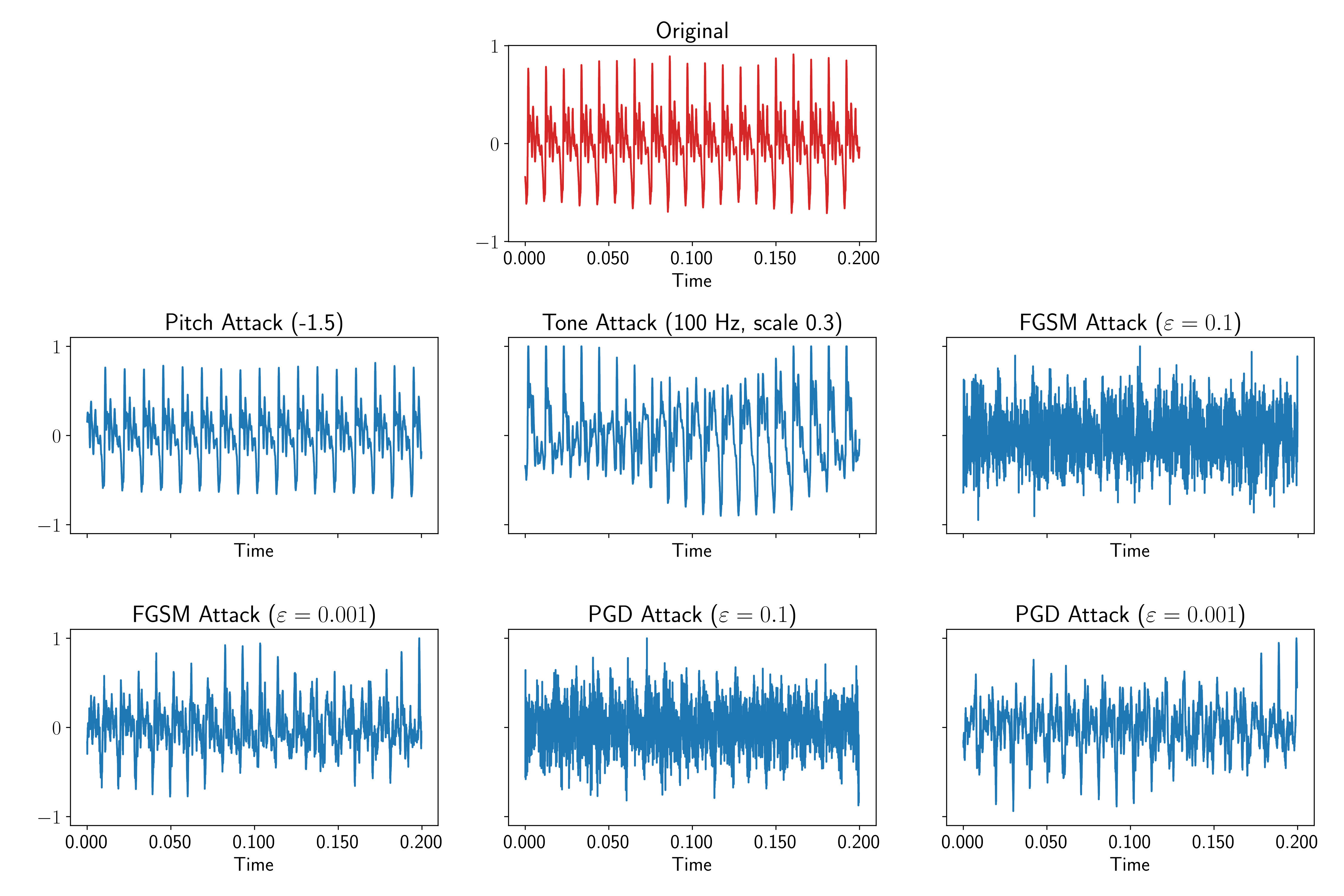}
    \caption{Audio snippet subjected to different evasion attacks, both black box (tone and pitching) and white box (PGD and FGSM).}
    \label{fig:visualaudio}
\end{figure}

In this paper, we analyzed the robustness of voice disorder detection systems to changes in input signals by evaluating both black-box and white-box attacks. Our results highlight a critical susceptibility of these systems to adversarial perturbations. To understand the extent of these perturbations on the original signals, we reported in Fig. \ref{fig:visualaudio} an example of original and manipulated signal for all attacks performed. While pitch and tone attacks introduce noticeable but specific alterations to the waveform, \textit{FGSM} and \textit{PGD} attacks, particularly at higher epsilon values, cause significant noise, demonstrating how these adversarial methods can distort the input signals and potentially compromise the detection systems.

The susceptibility of voice disorder detection systems to adversarial attacks poses a significant challenge to their reliability and effectiveness. Our future work will focus on developing robust defense mechanisms to enhance the resilience of these systems against such perturbations. A risk model will be associated with evaluating the concrete impact of neglecting true positives or caring for false positives with further medical treatments. In addition, the erroneous classification of normal audio as pathological can have significant ramifications outside clinical settings, as the case of cyberbullying or harassment campaigns. 
\section{Acknowledgements}
This work is supported by the European Union - NextGenerationEU within the PRIN 2022 PNRR - BullyBuster 2 – the ongoing fight against bullying and cyberbullying with the help of artificial intelligence for the human wellbeing (CUP: F53C2200074007, Proj. Code: P2022K39K8) and within the SERICS (PE00000014) under the Italian Ministry of University (MUR) and Research National Recovery and Resilience Plan.


\bibliographystyle{IEEEtran}
\bibliography{refs}

\begin{thebibliography}{10}
\providecommand{\url}[1]{#1}
\csname url@samestyle\endcsname
\providecommand{\newblock}{\relax}
\providecommand{\bibinfo}[2]{#2}
\providecommand{\BIBentrySTDinterwordspacing}{\spaceskip=0pt\relax}
\providecommand{\BIBentryALTinterwordstretchfactor}{4}
\providecommand{\BIBentryALTinterwordspacing}{\spaceskip=\fontdimen2\font plus
\BIBentryALTinterwordstretchfactor\fontdimen3\font minus \fontdimen4\font\relax}
\providecommand{\BIBforeignlanguage}[2]{{%
\expandafter\ifx\csname l@#1\endcsname\relax
\typeout{** WARNING: IEEEtran.bst: No hyphenation pattern has been}%
\typeout{** loaded for the language `#1'. Using the pattern for}%
\typeout{** the default language instead.}%
\else
\language=\csname l@#1\endcsname
\fi
#2}}
\providecommand{\BIBdecl}{\relax}
\BIBdecl

\bibitem{boone2005voice}
D.~Boone, ``The voice and voice therapy,'' \emph{Allyn and Bacon google schola}, vol.~2, pp. 830--843, 2005.

\bibitem{huston2024prevalence}
M.~N. Huston, I.~Puka, and M.~R. Naunheim, ``Prevalence of voice disorders in the united states: a national survey,'' \emph{The Laryngoscope}, vol. 134, no.~1, pp. 347--352, 2024.

\bibitem{umeno2020summary}
H.~Umeno, M.~Hyodo, T.~Haji, H.~Hara, M.~Imaizumi, M.~Ishige, M.~Kumada, K.~Makiyama, N.~Nishizawa, K.~Saito \emph{et~al.}, ``A summary of the clinical practice guideline for the diagnosis and management of voice disorders, 2018 in japan,'' \emph{Auris Nasus Larynx}, vol.~47, no.~1, pp. 7--17, 2020.

\bibitem{cohen2012direct}
S.~M. Cohen, J.~Kim, N.~Roy, C.~Asche, and M.~Courey, ``Direct health care costs of laryngeal diseases and disorders,'' \emph{The Laryngoscope}, vol. 122, no.~7, pp. 1582--1588, 2012.

\bibitem{paul2013diagnostic}
B.~C. Paul, S.~Chen, S.~Sridharan, Y.~Fang, M.~R. Amin, and R.~C. Branski, ``Diagnostic accuracy of history, laryngoscopy, and stroboscopy,'' \emph{The Laryngoscope}, vol. 123, no.~1, pp. 215--219, 2013.

\bibitem{alanazi2022using}
A.~Alanazi, ``Using machine learning for healthcare challenges and opportunities,'' \emph{Informatics in Medicine Unlocked}, vol.~30, p. 100924, 2022.

\bibitem{alotaibi2019implementation}
F.~S. Alotaibi, ``Implementation of machine learning model to predict heart failure disease,'' \emph{International Journal of Advanced Computer Science and Applications}, vol.~10, no.~6, 2019.

\bibitem{gupta2024voice}
R.~Gupta, D.~R. Gunjawate, D.~D. Nguyen, C.~Jin, and C.~Madill, ``Voice disorder recognition using machine learning: a scoping review protocol,'' \emph{BMJ open}, vol.~14, no.~2, p. e076998, 2024.

\bibitem{carlini2018audio}
N.~Carlini and D.~Wagner, ``Audio adversarial examples: Targeted attacks on speech-to-text,'' in \emph{2018 IEEE security and privacy workshops (SPW)}.\hskip 1em plus 0.5em minus 0.4em\relax IEEE, 2018, pp. 1--7.

\bibitem{xu2020adversarial}
H.~Xu, Y.~Ma, H.-C. Liu, D.~Deb, H.~Liu, J.-L. Tang, and A.~K. Jain, ``Adversarial attacks and defenses in images, graphs and text: A review,'' \emph{International journal of automation and computing}, vol.~17, pp. 151--178, 2020.

\bibitem{zhang2022deepfake}
T.~Zhang, ``Deepfake generation and detection, a survey,'' \emph{Multimedia Tools and Applications}, vol.~81, no.~5, pp. 6259--6276, 2022.

\bibitem{eskidere2015voice}
{\"O}.~Eskidere, A.~G{\"u}rhanl{\i} \emph{et~al.}, ``Voice disorder classification based on multitaper mel frequency cepstral coefficients features,'' \emph{Computational and mathematical methods in medicine}, vol. 2015, 2015.

\bibitem{al2017investigation}
A.~Al-Nasheri, G.~Muhammad, M.~Alsulaiman, Z.~Ali, T.~A. Mesallam, M.~Farahat, K.~H. Malki, and M.~A. Bencherif, ``An investigation of multidimensional voice program parameters in three different databases for voice pathology detection and classification,'' \emph{Journal of Voice}, vol.~31, no.~1, pp. 113--e9, 2017.

\bibitem{idrisoglu2023applied}
A.~Idrisoglu, A.~L. Dallora, P.~Anderberg, and J.~S. Berglund, ``Applied machine learning techniques to diagnose voice-affecting conditions and disorders: systematic literature review,'' \emph{Journal of Medical Internet Research}, vol.~25, p. e46105, 2023.

\bibitem{al2021voice}
F.~T. Al-Dhief, M.~M. Baki, N.~M.~A. Latiff, N.~N. N.~A. Malik, N.~S. Salim, M.~A.~A. Albader, N.~M. Mahyuddin, and M.~A. Mohammed, ``Voice pathology detection and classification by adopting online sequential extreme learning machine,'' \emph{IEEE Access}, vol.~9, pp. 77\,293--77\,306, 2021.

\bibitem{souissi2015dimensionality}
N.~Souissi and A.~Cherif, ``Dimensionality reduction for voice disorders identification system based on mel frequency cepstral coefficients and support vector machine,'' in \emph{2015 7th international conference on modelling, identification and control (ICMIC)}.\hskip 1em plus 0.5em minus 0.4em\relax IEEE, 2015, pp. 1--6.

\bibitem{teixeira2017vocal}
J.~P. Teixeira, P.~O. Fernandes, and N.~Alves, ``Vocal acoustic analysis--classification of dysphonic voices with artificial neural networks,'' \emph{Procedia computer science}, vol. 121, pp. 19--26, 2017.

\bibitem{benhammoud2018automatic}
R.~Benhammoud and A.~Kacha, ``Automatic classification of disordered voices with hidden markov models,'' in \emph{2018 International Conference on Signal, Image, Vision and their Applications (SIVA)}.\hskip 1em plus 0.5em minus 0.4em\relax IEEE, 2018, pp. 1--6.

\bibitem{ali2017automatic}
Z.~Ali, G.~Muhammad, and M.~F. Alhamid, ``An automatic health monitoring system for patients suffering from voice complications in smart cities,'' \emph{IEEE Access}, vol.~5, pp. 3900--3908, 2017.

\bibitem{reid2022development}
J.~Reid, P.~Parmar, T.~Lund, D.~K. Aalto, and C.~C. Jeffery, ``Development of a machine-learning based voice disorder screening tool,'' \emph{American Journal of Otolaryngology}, vol.~43, no.~2, p. 103327, 2022.

\bibitem{powell2019decoding}
M.~E. Powell, M.~Rodriguez~Cancio, D.~Young, W.~Nock, B.~Abdelmessih, A.~Zeller, I.~Perez~Morales, P.~Zhang, C.~G. Garrett, D.~Schmidt \emph{et~al.}, ``Decoding phonation with artificial intelligence (dep ai): proof of concept,'' \emph{Laryngoscope Investigative Otolaryngology}, vol.~4, no.~3, pp. 328--334, 2019.

\bibitem{fang2019detection}
S.-H. Fang, Y.~Tsao, M.-J. Hsiao, J.-Y. Chen, Y.-H. Lai, F.-C. Lin, and C.-T. Wang, ``Detection of pathological voice using cepstrum vectors: A deep learning approach,'' \emph{Journal of Voice}, vol.~33, no.~5, pp. 634--641, 2019.

\bibitem{khamaiseh2022adversarial}
S.~Y. Khamaiseh, D.~Bagagem, A.~Al-Alaj, M.~Mancino, and H.~W. Alomari, ``Adversarial deep learning: A survey on adversarial attacks and defense mechanisms on image classification,'' \emph{IEEE Access}, 2022.

\bibitem{liu2019adversarial}
S.~Liu, H.~Wu, H.-y. Lee, and H.~Meng, ``Adversarial attacks on spoofing countermeasures of automatic speaker verification,'' in \emph{2019 IEEE Automatic Speech Recognition and Understanding Workshop (ASRU)}.\hskip 1em plus 0.5em minus 0.4em\relax IEEE, 2019, pp. 312--319.

\bibitem{szegedy2013intriguing}
C.~Szegedy, W.~Zaremba, I.~Sutskever, J.~Bruna, D.~Erhan, I.~Goodfellow, and R.~Fergus, ``Intriguing properties of neural networks,'' \emph{arXiv preprint arXiv:1312.6199}, 2013.

\bibitem{madry2019deep}
A.~Madry, A.~Makelov, L.~Schmidt, D.~Tsipras, and A.~Vladu, ``Towards deep learning models resistant to adversarial attacks,'' 2019.

\bibitem{papernot2017practical}
N.~Papernot, P.~McDaniel, I.~Goodfellow, S.~Jha, Z.~B. Celik, and A.~Swami, ``Practical black-box attacks against machine learning,'' in \emph{Proceedings of the 2017 ACM on Asia conference on computer and communications security}, 2017, pp. 506--519.

\bibitem{ge2023advddos}
Y.~Ge, L.~Zhao, Q.~Wang, Y.~Duan, and M.~Du, ``Advddos: Zero-query adversarial attacks against commercial speech recognition systems,'' \emph{IEEE Transactions on Information Forensics and Security}, 2023.

\bibitem{zheng2021black}
B.~Zheng, P.~Jiang, Q.~Wang, Q.~Li, C.~Shen, C.~Wang, Y.~Ge, Q.~Teng, and S.~Zhang, ``Black-box adversarial attacks on commercial speech platforms with minimal information,'' in \emph{Proceedings of the 2021 ACM SIGSAC Conference on Computer and Communications Security}, 2021, pp. 86--107.

\bibitem{patil2023review}
P.~Patil and K.~Patil, ``A review on disease prediction using image processing,'' \emph{Journal Electrical and Computer Experiences}, vol.~1, no.~1, pp. 18--28, 2023.

\bibitem{app12168129}
\BIBentryALTinterwordspacing
Y.~Zhang, J.~Qian, X.~Zhang, Y.~Xu, and Z.~Tao, ``Pathological voice detection using joint subsapce transfer learning,'' \emph{Applied Sciences}, vol.~12, no.~16, 2022. [Online]. Available: \url{https://www.mdpi.com/2076-3417/12/16/8129}
\BIBentrySTDinterwordspacing

\bibitem{guan2019evaluation}
H.~Guan and A.~Lerch, ``Evaluation of feature learning methods for voice disorder detection,'' \emph{International Journal of Semantic Computing}, vol.~13, no.~04, pp. 453--470, 2019.

\bibitem{howard2019searching}
A.~Howard, M.~Sandler, G.~Chu, L.-C. Chen, B.~Chen, M.~Tan, W.~Wang, Y.~Zhu, R.~Pang, V.~Vasudevan \emph{et~al.}, ``Searching for mobilenetv3,'' in \emph{Proceedings of the IEEE/CVF international conference on computer vision}, 2019, pp. 1314--1324.

\bibitem{yousif2023generic}
N.~R. Yousif, H.~M. Balaha, A.~Y. Haikal, and E.~M. El-Gendy, ``A generic optimization and learning framework for parkinson disease via speech and handwritten records,'' \emph{Journal of Ambient Intelligence and Humanized Computing}, vol.~14, no.~8, pp. 10\,673--10\,693, 2023.

\end{thebibliography}
\end{document}